\documentclass{ifacconf}

\usepackage{amsmath}
\usepackage{amssymb}
\usepackage{booktabs}
\usepackage{graphicx}
\usepackage{natbib}
\usepackage{url}

\begin{document}
\begin{frontmatter}

\title{pysib: An Open-Source Python Toolbox for Linear System Identification}

\author[ufrgs]{Diego Eckhard}
\address[ufrgs]{Universidade Federal do Rio Grande do Sul, Porto Alegre, Brazil
  (e-mail: diegoeck@ufrgs.br)}

\begin{abstract}
Discrete-time polynomial input--output models (ARX, ARMAX, OE, and
Box--Jenkins) are usually estimated by prediction-error methods, but for OE,
ARMAX, and BJ the finite-sample criterion is nonconvex: the estimate a user
actually obtains is set by the initialization and the optimization procedure,
not only by the asymptotic theory.  This article documents the dedicated
optimization strategy behind \texttt{pysib}, an open-source Python toolbox for
SISO polynomial system identification.  The strategy consists of an ARX-based
initialization, a smoothed-gradient phase, an incremental Gauss--Newton
refinement, and filtered continuation interpreted as cost-function shaping.  This
strategy produced the
filtered-continuation results reported by the author in earlier work but had not
previously been described or released; it is given here in full and as open
Python software, with a common five-polynomial representation, shared prediction
and simulation routines, and the scripts and archived release needed to
reproduce the experiments.  On a moderate-noise OE benchmark the strategy returns
estimates far more concentrated around the true parameters than a
general-purpose nonlinear-programming solver, and on a harder nonconvex benchmark
filtered continuation raises the success rate from 60\% to 100\%.
\end{abstract}


\end{frontmatter}

\section{Introduction}
\label{sec:introduction}

Discrete-time polynomial input--output models remain a standard representation
for linear system identification.  In SISO applications, ARX, ARMAX, OE, and
Box--Jenkins structures provide compact transfer-function descriptions in which
the plant dynamics, input delay, and disturbance model can be represented by a
small number of polynomials.  This representation is attractive not only because
of its long history in system identification, but also because it gives models
that are directly usable for prediction, simulation, and control-oriented
analysis \citep{ljung2010}.

Prediction-error methods are commonly used to estimate these polynomial
structures from input--output data.  Under standard assumptions, including
sufficient excitation, identifiability, stability, and model correctness, the
global minimizer of the prediction-error criterion has the usual consistency and
asymptotic covariance properties \citep{ljung1999,soderstrom1989}.  These
properties, however, describe the ideal minimizer of the criterion.  They do not
by themselves specify how a finite-data, generally nonconvex optimization
problem should be solved in software.  For ARX this objective is linear least
squares, but for OE, ARMAX, and BJ structures it is generally nonconvex because
denominator and disturbance-model parameters enter the prediction recursion.
Consequently, the numerical initialization and local optimization strategy
become central to the estimator that the procedure yields in practice.

These numerical issues motivate software in which the model convention and the
optimization procedure are both explicit.  A system-identification package
should use a consistent representation across estimators and should make it
possible to reproduce the numerical evidence supporting its use.  \texttt{pysib}
is an open-source Python toolbox for SISO polynomial system identification
\citep{pysib2026}.  Its scope is deliberately narrow: the package addresses
discrete-time SISO polynomial input--output models, rather than MIMO
identification, subspace methods, continuous-time identification, or nonlinear
machine-learning models \citep{chen2012}.  Within this scope, it provides direct estimators,
nonlinear PEM estimators, filtered variants, and a common prediction and
simulation interface.  This focus makes it possible to expose the polynomial
structure used by the estimators while keeping the public interface compact.

Open Python tooling for this classical polynomial PEM setting is comparatively
scarce, and the existing packages mostly address different problems.
\texttt{SysIdentPy} targets NARMAX and other nonlinear model classes, with model
structure selection and least-squares-type parameter estimation rather than the
nonconvex transfer-function PEM considered here \citep{sysidentpy}.
General-purpose control libraries such as \texttt{python-control} focus on
analysis and design and offer only limited identification functionality
\citep{pythoncontrol}.  The closest comparator is SIPPY, which does implement the
ARX, ARMAX, OE, and BJ structures, but casts the polynomial criteria as a general
nonlinear program solved with an off-the-shelf optimizer \citep{sippy}.  These
tools are complementary rather than directly comparable; \texttt{pysib} occupies
the specific niche of SISO polynomial PEM and pairs the polynomial convention
with an optimization procedure designed around the structure of the PEM
criterion, which is the part of the toolbox examined in most detail below.

The contribution of this article is threefold.  First, it gives the first
explicit description of the dedicated numerical strategy used by \texttt{pysib}
for nonlinear PEM estimation in OE, ARMAX, and BJ structures.  The strategy
combines ARX-based initialization, a smoothed gradient phase, a Gauss--Newton
refinement, and filtered continuation interpreted as cost-function shaping, in
which the filters guide the optimizer through intermediate filtered criteria
before returning to the original prediction-error problem.  This strategy
underlies the filtered-continuation results previously reported by the author
\citep{eckhard2013input,eckhard2017cost}, where it was used to produce
comparisons against established identification toolboxes but was never itself
documented.  Second, the article provides an open-source Python implementation
of the strategy, with a consistent polynomial model convention and a common
prediction and simulation interface, addressing the scarcity of Python tooling
for classical polynomial PEM.  Third, it reports reproducible OE experiments that
isolate the optimizer against a general-purpose solver and evaluate filtered
continuation on a nonconvex benchmark.  No new PEM theory is claimed; the
emphasis is on disclosing and documenting an effective optimization strategy,
releasing it as open software, and providing reproducible numerical evidence for
the tested SISO OE settings.

The remainder of the article is organized as follows.  Section~\ref{sec:polynomial-siso-models}
reviews the polynomial SISO model convention and PEM criterion.
Section~\ref{sec:optimization-initialization} describes the initialization,
optimization, and filtered-continuation strategy.  Section~\ref{sec:software-reproducibility}
summarizes the software interface and reproducibility information.
Section~\ref{sec:numerical-experiments} presents the numerical experiments, and
the final section gives concluding remarks.

\section{Prediction-Error Identification in Polynomial SISO Models}
\label{sec:polynomial-siso-models}

Consider a discrete-time SISO data record
\(\{u(t),y(t)\}_{t=1}^{N}\).  The objective is to estimate a finite-dimensional
model of the relation between the measured input \(u(t)\) and the measured
output \(y(t)\).  It is useful to distinguish the deterministic response of the
plant from the stochastic component of the measurement.  The noise-free output
is denoted by \(y_0(t)\), and the measured output is written as the sum of this
component and an additive disturbance:
\begin{align}
  y_0(t) &= G_0(q^{-1})u(t), \\
  v(t)   &= H_0(q^{-1})e(t), \\
  y(t)   &= y_0(t) + v(t).
  \label{eq:true-system}
\end{align}
Here \(q^{-1}\) is the backward-shift operator, \(G_0\) is the true plant,
\(H_0\) is the disturbance model, and \(e(t)\) denotes the innovation process in
the standard prediction-error setting.  This notation separates the dynamics to
be reproduced in simulation, represented by \(G_0\), from the dynamics used to
describe the color of the disturbance, represented by \(H_0\).

The model family implemented in \texttt{pysib} follows the classical
five-polynomial SISO convention \citep{ljung1999,soderstrom1989}.  This
convention is useful because several standard structures can be expressed by
fixing selected polynomials to unity and estimating the remaining ones.  With
parameter vector \(\theta\) and input delay \(n_k\), the candidate model is
represented as
\begin{equation}
  y(t;\theta) = G(q^{-1};\theta)u(t-n_k)
  + H(q^{-1};\theta)e(t),
  \label{eq:model-family}
\end{equation}
where
\begin{align}
  G(q^{-1};\theta)
  &= \frac{B(q^{-1};\theta)}{A(q^{-1};\theta)F(q^{-1};\theta)}, \\
  H(q^{-1};\theta)
  &= \frac{C(q^{-1};\theta)}{A(q^{-1};\theta)D(q^{-1};\theta)}.
  \label{eq:gh}
\end{align}
The polynomials are parameterized as
\begin{align}
  A(q^{-1}) &= 1+a_1q^{-1}+\cdots+a_{n_a}q^{-n_a}, \\
  B(q^{-1}) &= b_1+b_2q^{-1}+\cdots+b_{n_b}q^{-(n_b-1)}, \\
  C(q^{-1}) &= 1+c_1q^{-1}+\cdots+c_{n_c}q^{-n_c}, \\
  D(q^{-1}) &= 1+d_1q^{-1}+\cdots+d_{n_d}q^{-n_d}, \\
  F(q^{-1}) &= 1+f_1q^{-1}+\cdots+f_{n_f}q^{-n_f}.
  \label{eq:polynomial-parametrization}
\end{align}
Thus, \(n_b\) denotes the number of numerator coefficients in \(B\).  In this
conceptual parametrization, the input delay is not included in \(B\); it is the
separate integer \(n_k\) appearing in \eqref{eq:model-family}.
The transfer function \(G(q^{-1};\theta)\) represents the deterministic
input--output dynamics, whereas \(H(q^{-1};\theta)\) represents the disturbance
model used by the one-step-ahead predictor.  In the model dictionary returned by
the software, the same delay is stored by prepending \(n_k\) zeros to the
polynomial \(B\).  This storage convention is used for simulation and filtering,
but the order parameter \(n_b\) still counts only the nonzero numerator
coefficients.  In summary, the conceptual model carries the delay explicitly via
\(n_k\) while \(B\) contains no delay; the software prepends \(n_k\) zeros to the
returned \(B\) and does not expose \(n_k\) again through the model dictionary.

For a given \(\theta\), the one-step-ahead predictor associated with the model
is the standard expression
\begin{equation}
  \begin{aligned}
  \hat y(t\mid t-1;\theta)= &
  H^{-1}(q^{-1};\theta)G(q^{-1};\theta)u(t-n_k) \\
  & +\bigl[1-H^{-1}(q^{-1};\theta)\bigr]y(t),
  \end{aligned}
  \label{eq:predictor}
\end{equation}
which separates the deterministic filtered-input contribution from the
autoregressive contribution of the past output.  The prediction error
and the finite-sample least-squares prediction-error criterion are
\begin{align}
  \varepsilon(t,\theta) &= y(t)-\hat y(t\mid t-1;\theta), \\
  V_N(\theta) &= \frac{1}{N}\sum_{t=1}^{N}\varepsilon^2(t,\theta).
  \label{eq:pem-criterion}
\end{align}
Thus, once a polynomial structure and model orders are selected, the estimation
problem for the nonlinear structures is
\begin{equation}
  \hat\theta_N = \arg\min_{\theta\in\Theta}\, V_N(\theta),
  \label{eq:pem-problem}
\end{equation}
where \(\Theta\) denotes the set of free coefficients of the chosen structure
(for OE, \(\theta=[b_1,\dots,b_{n_b},f_1,\dots,f_{n_f}]\) with the constraint
that the roots of the estimated polynomial \(F\) lie inside the unit circle).
The nonlinear estimators in this article are numerical procedures for solving
\eqref{eq:pem-problem}.

The motivation for using \eqref{eq:pem-criterion} is the usual asymptotic
theory of prediction-error identification: under standard assumptions on
identifiability, excitation, stability, and model correctness, the global
minimizer has the classical consistency properties as \(N\to\infty\), and a
correctly parameterized disturbance model yields the standard asymptotic
efficiency result \citep{ljung1999,soderstrom1989,ljung1978}.  These statements concern
the global minimizer and the large-sample limit; they do not imply that the
criterion obtained from a finite data record is easy to minimize.

\begin{table}[t]
  \caption{Polynomial structures supported by \texttt{pysib}.}
  \label{tab:structures}
  \centering
  \footnotesize
  \begin{tabular}{lccccc}
    \toprule
    Structure & \(A\) & \(B\) & \(C\) & \(D\) & \(F\) \\
    \midrule
    ARX   & free & free & 1    & 1    & 1 \\
    ARMAX & free & free & free & 1    & 1 \\
    OE    & 1    & free & 1    & 1    & free \\
    BJ    & 1    & free & free & free & free \\
    \bottomrule
  \end{tabular}
\end{table}

Table~\ref{tab:structures} summarizes the polynomial structures considered in
the package.  ARX fixes the noise model and is linear in the estimated
parameters, so it can be computed by least squares and also serves as a natural
direct estimate for initialization.  OE fixes the disturbance model to white
noise and estimates the plant numerator and denominator, which is appropriate
when the main interest is the deterministic input--output dynamics.  ARMAX adds
a moving-average disturbance polynomial to the ARX structure, and BJ separates
the plant and disturbance denominators, allowing a more flexible noise model.

These structural choices have direct numerical consequences.  ARX leads to a
linear least-squares problem.  In contrast, OE, ARMAX, and BJ contain denominator
and/or disturbance-model parameters inside the prediction recursion, so the
resulting PEM criteria are generally nonconvex
\citep{astrom1974,soderstrom1982}.  Their practical use therefore
depends not only on writing down \eqref{eq:pem-criterion}, but also on computing
a useful local minimizer from finite data.  The dedicated initialization and
numerical optimization strategy used for this purpose is the subject of the next
section.

\section{Numerical Optimization and Initialization Strategy}
\label{sec:optimization-initialization}

For the nonlinear structures in Section~\ref{sec:polynomial-siso-models}, the
numerical procedure is an essential part of the practical estimator.  The PEM
criterion is defined by \eqref{eq:pem-criterion}, but for OE, ARMAX, and BJ the
finite-sample optimization problem is nonconvex and the result may depend on
the initial point and on the path followed by the optimizer.  A software
implementation must therefore make concrete choices about initialization,
descent safeguards, curvature information, and failure handling.  Local
convergence of maximum-likelihood and prediction-error iterations for
ARMAX-type models is known to depend on problem-specific conditions
\citep{goodwin2003}; the aim here is therefore not a new global-convergence
guarantee, but a reliable and reproducible local-search strategy designed around
the structure of the PEM criterion for the SISO polynomial models considered by
the toolbox.

A single principle guides these choices.  On the nonconvex PEM criteria of
interest, aggressive optimizers tend to fail in two ways: they settle into poor
local minima, or they step outside the stability region, most often by driving
the roots of \(F\) outside the unit circle, where the predictor diverges and the
cost becomes non-finite, the latter being the more frequent failure in practice.
The strategy adopted here deliberately trades iteration speed for reliability:
many small, smoothed steps and a gradually increasing Gauss--Newton step keep the
iterates from overshooting into either failure.  Stability is not tested
directly; an unstable trial point simply raises the cost or makes it non-finite,
so a single cost-based safeguard handles both failure modes at once by rejecting
and backtracking any trial point that does not decrease the cost.  The price
is a large iteration count, but the inner recursions are implemented in C, so the
method stays fast in wall-clock time.  This ``slow is better'' principle gives the
toolbox part of its name: SIB denotes both Slow Is Better and System
Identification Toolbox, and \texttt{pysib} is its Python implementation.

The first design choice is to start each nonlinear search from an auxiliary ARX
estimate.  This gives a deterministic and inexpensive point in the parameter
space before the nonlinear PEM criterion is optimized.  Although the ARX model
does not generally have the same disturbance structure as OE, ARMAX, or BJ, it
provides a direct polynomial approximation with numerator and denominator
dimensions compatible with the requested structure.  The resulting coefficients
initialize the corresponding plant polynomials, while the remaining polynomial
coefficients are set to neutral initial values.  Thus, the nonlinear stage
starts from a model that is dimensionally compatible with the requested
structure, rather than from an arbitrary parameter vector.

Starting from this point, the optimizer proceeds in two phases.  The reason for
using two phases is that the information useful near a local minimizer is not
necessarily reliable at the beginning of the search.  A Gauss--Newton step is
attractive when the current iterate is already in a region where the local
least-squares approximation is informative, but it can be fragile when the
initial estimate is still far from such a region or when the sensitivity matrix
is ill conditioned.  In preliminary testing, a direct damped Gauss--Newton or
Levenberg--Marquardt iteration applied to the nonlinear PEM criterion was found
unreliable under exactly these conditions, which motivated the two-phase strategy
adopted here.  The first phase is therefore deliberately conservative: it
follows a smoothed gradient direction and uses small adaptive changes in the step
length.  This phase is intended to obtain a decrease of the criterion without
first assuming that a local quadratic approximation is accurate.  Let
\(g_k = \nabla V_N(\theta_k)\) denote the gradient of the criterion
\eqref{eq:pem-criterion} evaluated at the current iterate.  Instead of using
\(g_k\) directly, the implementation updates a filtered direction
\begin{equation}
  d_k = \frac{4d_{k-1}+g_k}{5},
  \label{eq:direction-filter}
\end{equation}
which damps abrupt changes induced by finite-data effects or by the recursive
calculation of prediction errors.  The smoothed direction is a conservative
local search direction: it keeps memory of the previous direction while still
responding to the current gradient.  The trial point is then computed along the
normalized direction,
\begin{equation}
  \theta_{\mathrm{trial}}
  = \theta_k - \alpha_k\frac{d_k}{\lVert d_k\rVert}.
  \label{eq:gradient-step}
\end{equation}
The normalization separates the choice of direction from the scalar step
length, so that the same step-length variable can be used even when the gradient
magnitude changes substantially across iterations.  If the trial point
increases the cost, the step length is reduced; otherwise the point is accepted
and the step length is increased:
\begin{equation}
  \alpha_k \leftarrow
  \begin{cases}
    0.99\,\alpha_k, & \text{if } V_N(\theta_{\mathrm{trial}}) > V_N(\theta_k), \\
    1.01\,\alpha_k, & \text{otherwise}.
  \end{cases}
  \label{eq:step-adapt}
\end{equation}
This rule is a numerical safeguard.  It
should be read as a practical descent heuristic, not as a global-convergence
claim for the nonconvex PEM criterion.  The notation
\(\theta_{\mathrm{trial}}\) is used because the point is only provisional: if it
is accepted, it becomes the next iterate \(\theta_{k+1}\); otherwise
\(\theta_k\) is retained and only the step length is reduced.

After this initial decrease phase, the optimizer switches to a local refinement
based on a Gauss--Newton approximation.  At this stage, the algorithm uses the
sensitivity of the prediction errors to parameter perturbations, which provides
more directional information than the gradient alone when the current iterate is
close enough to a meaningful basin of attraction.  Let
\(S(\theta)\in\mathbb{R}^{N\times n_\theta}\) be the sensitivity matrix of the
prediction errors with respect to the free parameters, with entries
\(S_{tj}(\theta)=\partial\varepsilon(t,\theta)/\partial\theta_j\).  Up to scalar
factors that do not affect the search direction, the Gauss--Newton curvature
approximation is
\begin{equation}
  H_{\mathrm{GN}}(\theta)=S(\theta)^\top S(\theta),
  \label{eq:gn-hessian}
\end{equation}
and the search direction is obtained from
\begin{equation}
  H_{\mathrm{GN}}(\theta_k)p_k = g_k.
  \label{eq:gn-system}
\end{equation}
When \(S(\theta_k)\) has full column rank, \(H_{\mathrm{GN}}(\theta_k)\) is
positive definite.  The system \eqref{eq:gn-system} is solved for \(p_k\), and
because it approximates \(H_{\mathrm{GN}}^{-1}g_k\), the Newton-like descent
direction is \(-p_k\); this is why the trial update shown next subtracts
\(\beta_k p_k\) rather than adding it.  Because this approximation is most
reliable only near a local minimizer, the implementation does not immediately
take a full step.  It instead forms trial points
\begin{equation}
  \theta_{\mathrm{trial}} = \theta_k - \beta_k p_k,
  \qquad \beta_k = \frac{k}{1000},\quad k=1,\ldots,1000,
  \label{eq:incremental-newton}
\end{equation}
so that the effective Gauss--Newton step increases gradually over the
iterations.  This gradual transition is intended to combine the robustness of
the previous descent phase with the faster local progress of curvature-based
updates.  In other words, the method does not abruptly replace the conservative
phase by an aggressive full Newton step; it introduces curvature information
progressively.  If a trial point increases the cost or produces a nonfinite
value, it is moved halfway back to the current iterate,
\begin{equation}
  \theta_{\mathrm{trial}}
  \leftarrow \frac{\theta_{\mathrm{trial}}+\theta_k}{2},
  \label{eq:bisection-backtracking}
\end{equation}
and the test is repeated a fixed number of times.  Once a trial point satisfies
the cost-decrease test, it is accepted as the next iterate.  The overall
procedure is
therefore a safeguarded local search: it first seeks a reliable decrease of the
criterion and then uses sensitivity information to refine the estimate.

Each phase runs for a fixed iteration budget rather than a gradient-norm
stopping test.  The smoothed-gradient phase performs at most \(10^{4}\) step
updates, organized as one hundred outer iterations of one hundred inner
step-length adaptations, and exits early when the adaptive step length collapses
below \(10^{-7}\).  The Gauss--Newton phase then runs a fixed budget of one
thousand iterations, over which the incremental factor \(\beta_k\) grows from
near zero toward a full step.  Fixing these budgets keeps the procedure
deterministic and reproducible for a given data record.

The same fixed configuration is used unchanged for every structure and every
experiment reported here: the smoothing weight, the step-length factors, the
Gauss--Newton ramp, and the iteration budgets are not tuned per problem.  This
deliberate design choice favours robustness and determinism over per-problem
tuning.  Because the experiment scripts and the random seed are provided, the
absence of tuning is auditable rather than merely asserted.

The computationally intensive part of this process is the repeated evaluation
of prediction errors, sensitivities, and gradients.  These quantities are
computed by sequential filtering recursions: each new sample depends on previous
values of the filtered signals and sensitivity variables.  Such recursions are
not naturally expressible as a small number of vectorized array operations.
For this reason, the inner recursions for OE, ARMAX, and BJ are implemented in
C.  This implementation choice keeps the recursion formulas close to the
polynomial model structure and avoids Python loops in the numerical core, while
the public estimator interface remains in Python.

The optimization procedure described so far acts within a single objective
function: the two phases change how the search moves, but the criterion
\(V_N(\theta)\) defined by \eqref{eq:pem-criterion} stays the same throughout
the call.  In practice, however, the finite-sample shape of \(V_N\) can itself
present the main difficulty, creating basins of attraction that lead a
well-designed local search toward unacceptable local minima irrespectively of
step-length choices.  The filtered estimators described next address this
problem at a different level: instead of modifying the optimizer, they modify
the sequence of criteria presented to it.

\subsection{Filtered Continuation as Cost-Function Shaping}
\label{subsec:filtered-continuation-shaping}

The filtered estimators add a second mechanism for reducing the dependence on a
single difficult finite-sample criterion.  The preceding paragraphs describe how
the optimizer moves within one criterion; filtered continuation acts at a
different level by changing the sequence of criteria presented to the optimizer.
It should not be interpreted merely as a preprocessing option applied before
identification.  If both input and output data are filtered and the PEM
criterion is evaluated on the filtered record, the optimizer sees a different
finite-sample objective.  Each filter therefore defines an auxiliary criterion
with its own landscape of local minima.  Changing the filter changes the
relative influence of different frequency regions and may make the early
optimization problems easier than the original unfiltered problem.

This interpretation is the cost-function-shaping view of filtered OE
identification \citep{eckhard2011global,eckhard2013input,eckhard2017cost}.  The purpose of the
filters is to guide the optimizer through a sequence of related criteria in
which undesirable local minima may have reduced influence.  The wording is
important: filtering can reshape the finite-sample criterion and can improve the
numerical path followed by a local optimizer, but it does not remove the
nonconvexity of the final PEM problem and does not provide a general guarantee
of reaching its global minimum.

In \texttt{pysib}, the filtered variants implement this idea as a continuation
procedure.  A filtered identification problem is solved first; its solution is
then used as the initial condition for the next, less restrictive filtered
problem.  The sequence continues until the final stage, which is the original
unfiltered PEM criterion.  Consequently, the returned estimate is still an
estimate for the original model structure and criterion; the filters are used to
shape the path toward that final problem.  In the OE, ARMAX, and BJ variants,
nine filtered stages are applied before the final unfiltered PEM call.  Each
stage uses a
first-order low-pass filter whose pole is computed as
\begin{equation}
  p = \exp\bigl(\log 0.05/(\tau\cdot 40)\bigr)
  \label{eq:oe-pole}
\end{equation}
with \(\tau=0.9,0.8,\ldots,0.1\); the
resulting poles range from approximately \(0.92\) to \(0.47\), progressively
relaxing the filtering constraint.  In all cases the continuation ends with
the original unfiltered PEM criterion.

This mechanism is directly examined in the numerical experiments of
Section~\ref{sec:numerical-experiments}.  There, the standard OE estimator and
the filtered OE continuation are applied to the same nonconvex benchmark, so
that the effect of solving the intermediate filtered problems can be evaluated
without changing the final simulation-error metric.

\section{Software Interface and Reproducibility}
\label{sec:software-reproducibility}

The public interface of \texttt{pysib} is organized around the SISO
input--output data record and the requested model orders.  Each estimator returns
a pair \((\hat\theta,m)\), where \(\hat\theta\) is the vector of free estimated
parameters and \(m\) is a structured polynomial representation of the identified
model.  This convention separates the numerical parameter vector used by the
optimizer from the model object used for prediction, simulation, and comparison.
The vector \(\hat\theta\) is useful when the user wants to inspect the numerical
parameters directly, for example in Monte Carlo studies of parameter dispersion.
The structured object \(m\), in contrast, is the representation used by the rest
of the package to evaluate the identified model.

The model representation \(m\) is a dictionary with keys \texttt{A}, \texttt{B},
\texttt{C}, \texttt{D}, and \texttt{F}.  These entries correspond directly to
the five polynomials introduced in Section~\ref{sec:polynomial-siso-models};
polynomials that are fixed by a particular structure are represented by the
scalar polynomial \([1]\).  The same convention is used by the linear estimators,
the nonlinear PEM estimators, and the filtered variants.  As a result, changing
the estimator does not require changing the downstream code used for prediction,
simulation, or error computation.

\begin{table}[t]
  \caption{Public routines and their roles.}
  \label{tab:public-routines}
  \centering
  \scriptsize
  \begin{tabular}{lll}
    \toprule
    Routine & Structure or method & Role \\
    \midrule
    \texttt{arx} & ARX & Linear/init. \\
    \texttt{sm} & Stieglitz--McBride & OE init. \\
    \texttt{iv} & Instrumental variables & ARX alternative \\
    \texttt{correlation} & Correlation & ARX alternative \\
    \texttt{oe} & OE & Nonlinear PEM \\
    \texttt{armax} & ARMAX & Nonlinear PEM \\
    \texttt{bj} & Box--Jenkins & Nonlinear PEM \\
    \texttt{*\_filtered} & Filtered continuation & Cost shaping \\
    \texttt{predict/simulate} & Model evaluation & Common evaluation \\
    \bottomrule
  \end{tabular}
\end{table}

Table~\ref{tab:public-routines} summarizes the main routines.  The first group
contains direct or linear methods that can be used either as estimators in their
own right or as auxiliary initial conditions.  The \texttt{sm} routine implements
the Stieglitz--McBride method \citep{stieglitz1965}.  The nonlinear routines
\texttt{oe}, \texttt{armax}, and \texttt{bj} call the PEM optimizer described in
Section~\ref{sec:optimization-initialization}.  The filtered variants expose the
continuation strategy of Section~\ref{subsec:filtered-continuation-shaping}
through the same model convention.  This common interface is also important for
the experiments in Section~\ref{sec:numerical-experiments}, because all methods
can be evaluated by the same simulation routine and the same error metric.

The direct methods \texttt{iv}, \texttt{correlation}, and \texttt{sm} do not
rely on a PEM optimizer and are included as diagnostic and initialization tools.
Instrumental variables reduce the bias introduced by colored output noise by
using auxiliary instruments, typically at the cost of increased variance compared
to least squares.  The correlation estimator provides an alternative ARX estimate
based on sample correlations rather than the normal equations.  The
Stieglitz--McBride method \citep{stieglitz1965} is a classical iterative
procedure for OE models that can diverge in some configurations but is often
effective as a fast initializer for nonlinear PEM.  All three estimators can be
used as stand-alone routines and their output models can also supply initial
parameter vectors for the \texttt{oe}, \texttt{armax}, and \texttt{bj} routines.

The following minimal example illustrates the common interface for OE
identification and model evaluation:
{\footnotesize\begin{verbatim}
import pysib

theta, m = pysib.oe(u, y, nb=1, nf=3, nz=1)
yp = pysib.predict(u, y, m)
ys = pysib.simulate(u, m)
\end{verbatim}}
Here \(u\) and \(y\) are the measured input and output vectors.  The call to
\texttt{oe} estimates an OE model with one numerator coefficient, a third-order
\(F\) polynomial, and one sample of input delay.  The same returned model
representation \(m\) is then used for one-step-ahead prediction and for
noise-free simulation.  The same pattern applies to models returned by the other
estimators: once \(m\) has been constructed, the evaluation code is independent
of the routine that produced it.

Any model returned by an estimator can be passed to the common \texttt{predict}
and \texttt{simulate} routines.  The former computes one-step-ahead predictions
using the model disturbance structure, whereas the latter computes the simulated
noise-free response of the plant model.  Keeping these operations separate is
important in system-identification experiments, since prediction and simulation
measure different aspects of the identified model.  This evaluation layer is
also the mechanism used to compute the simulation-error metric in the numerical
experiments of Section~\ref{sec:numerical-experiments}.

Formally, given a model \(m\) returned by any estimator, the \texttt{predict}
routine evaluates the one-step-ahead predictor \eqref{eq:predictor} using the
polynomials in \(m\), with \(G\) and \(H\) formed from those polynomials and the
input delay included in \(B\) according to the storage convention.  The
noise-free simulation computed by \texttt{simulate} uses only the plant model,
\begin{equation}
  \hat y_s(t) = G(q^{-1})u(t).
  \label{eq:simulate-routine}
\end{equation}
Thus, \texttt{predict} uses the full model (plant and disturbance), while
\texttt{simulate} uses only the deterministic plant dynamics.

Table~\ref{tab:availability} reports the software version and access points used
for the artifact evaluated in this article.

\begin{table}[ht]
  \caption{Software availability.}
  \label{tab:availability}
  \centering
  \footnotesize
  \begin{tabular}{lp{0.55\columnwidth}}
    \toprule
    Version & \texttt{pysib} v0.2.3~\citep{pysib2026} \\
    License & MIT \\
    Install & \texttt{pip install pysib} \\
    Documentation & \url{https://pysib.net/} \\
    Source & \url{https://github.com/diegoeck/pysib} \\
    DOI & \url{https://doi.org/10.5281/zenodo.20572074} \\
    Tests and experiments & Included in repository \\
    \bottomrule
  \end{tabular}
\end{table}

Detailed API documentation is left to the accompanying
software documentation; the purpose of this article is to describe the numerical
methods and the evidence obtained from the reproducible experiments.

\section{Numerical Experiments}
\label{sec:numerical-experiments}

The numerical experiments are restricted to SISO OE examples.  This choice keeps
the empirical analysis focused on plant-dynamics estimation and on the numerical
behavior of the PEM optimizer, without adding the separate issue of estimating a
colored-noise model.  OE is also a direct representative of the nonconvex
polynomial PEM problems targeted by the dedicated optimization strategy.  The two
experiments are designed to probe that strategy: the first contrasts the
dedicated PEM optimizer with a general-purpose solver on a benign problem, and
the second isolates the effect of filtered continuation on a harder nonconvex
problem.  Together they provide empirical evidence for the design choices of
Section~\ref{sec:optimization-initialization} in the tested OE settings.  Broader
comparisons of the filtered method against established MATLAB identification
toolboxes are reported in earlier work \citep{eckhard2013input,eckhard2017cost};
the present experiments instead isolate the optimizer and the continuation within
a single reproducible toolbox.  The
ARMAX and BJ structures use the same optimization strategy and interface;
they have been exercised during package development, but a systematic Monte
Carlo evaluation comparable to the OE experiments is left for future work.

Both experiments use the same third-order OE plant.  The conceptual
parametrization of Section~\ref{sec:polynomial-siso-models} gives
\begin{equation}
  \begin{aligned}
    B(q^{-1}) &= 1, \quad n_k = 1, \\
    F(q^{-1}) &= 1-2.4q^{-1}+1.91q^{-2}-0.504q^{-3}.
  \end{aligned}
  \label{eq:experiment-plant}
\end{equation}
For each experiment, the noise-free output \(y_0\) is generated by simulating
\eqref{eq:experiment-plant} with the specified input sequence, and the measured
output is
\begin{align}
  y(t) &= y_0(t)+v(t), \\
  v(t) &\sim \mathcal{N}(0,\sigma_v^2).
  \label{eq:experiment-noise}
\end{align}
The sample index is \(t=1,\ldots,N\), and the random seed is fixed to zero in
the experiment scripts.
The main metric is the relative simulation error
\begin{equation}
  E_{\mathrm{sim}} = \frac{\lVert y_0 - \hat y\rVert_2}{\lVert y_0\rVert_2}.
  \label{eq:simulation-error}
\end{equation}
Here \(y_0\) is the noise-free plant output and \(\hat y\) is the simulated
output of the estimated model.  This metric evaluates the estimated plant
dynamics rather than the one-step-ahead predictor, which is appropriate for the
OE benchmarks considered here.  A run is counted as successful when
\(E_{\mathrm{sim}}<5\%\).  This threshold is used only as an operational marker
of a low simulation error, not as a theoretical consistency criterion.

\subsection{Noisy-Data Comparison with SIPPY}
\label{subsec:noisy-data-sippy}

The first experiment compares \texttt{pysib} with the SIPPY OE implementation
\citep{sippy} (the \texttt{sippy\_unipi} distribution, version 1.0.1, called
through its OE system-identification routine with matching model orders), using
noisy data generated from
\eqref{eq:experiment-plant}.  The two implementations take different routes to
the same OE structure: SIPPY formulates the problem as a general nonlinear
program solved with an off-the-shelf nonlinear optimizer, whereas \texttt{pysib}
applies the dedicated PEM strategy of
Section~\ref{sec:optimization-initialization} from an ARX initialization.  The
comparison is therefore between a general-purpose solver and an optimizer built
around the PEM criterion.  The noise level
\(\sigma_v=1\) defines a moderate-noise setting in which all methods are expected
to identify the plant reasonably well, so the relevant quantities are the
concentration of the estimated parameters and the distribution of the simulation
error.  In other words, this benchmark does not test whether the plant is
identifiable, but how tightly the underlying optimization converges to the
data-generating parameters in a benign, near-convex case.  The Monte Carlo study uses \(M=100\) independent data records of length
\(N=1000\).  Additive white output noise is added to the noise-free output.  The
input is the deterministic multisine
\begin{equation}
  u(t)=\sin(2\pi t/18)+\sin(2\pi t/28)+\sin(2\pi t/61).
  \label{eq:noisy-input}
\end{equation}
The methods compared are the Stieglitz--McBride estimator in \texttt{pysib}, the
OE PEM estimator in \texttt{pysib}, and the SIPPY OE estimator, all using the
same OE order.  The parameter table is included to show whether the estimates
are centered near the coefficients of the data-generating plant, while the error
table and figure show the effect of these estimates on simulated output.

\begin{table}[ht]
  \caption{Parameter statistics for the noisy-data experiment.}
  \label{tab:noisy-parameter}
  \centering
  \scriptsize
  \begin{tabular}{llcc}
    \toprule
    Method & Parameter & Mean & Std \\
    \midrule
    SM & \(b_1\) & 0.9997 & 0.0020 \\
    SM & \(f_1\) & -2.4002 & 0.0012 \\
    SM & \(f_2\) & 1.9103 & 0.0023 \\
    SM & \(f_3\) & -0.5041 & 0.0011 \\
    OE & \(b_1\) & 0.9997 & 0.0020 \\
    OE & \(f_1\) & -2.4002 & 0.0012 \\
    OE & \(f_2\) & 1.9103 & 0.0023 \\
    OE & \(f_3\) & -0.5041 & 0.0011 \\
    SIPPY & \(b_1\) & 1.0203 & 0.0298 \\
    SIPPY & \(f_1\) & -2.3870 & 0.0168 \\
    SIPPY & \(f_2\) & 1.8860 & 0.0311 \\
    SIPPY & \(f_3\) & -0.4929 & 0.0145 \\
    \bottomrule
  \end{tabular}
\end{table}

\begin{table}[ht]
  \caption{Simulation-error statistics for the noisy-data experiment.}
  \label{tab:noisy-error}
  \centering
  \scriptsize
  \begin{tabular}{lccc}
    \toprule
    Method & Mean error & Std error & Success \\
    \midrule
    SM & \(7.28\times 10^{-4}\) & \(2.64\times 10^{-4}\) & 100\% \\
    OE & \(7.26\times 10^{-4}\) & \(2.65\times 10^{-4}\) & 100\% \\
    SIPPY & \(6.65\times 10^{-3}\) & \(5.49\times 10^{-3}\) & 100\% \\
    \bottomrule
  \end{tabular}
\end{table}

\begin{figure}[htb]
  \begin{center}
    \includegraphics[width=\columnwidth]{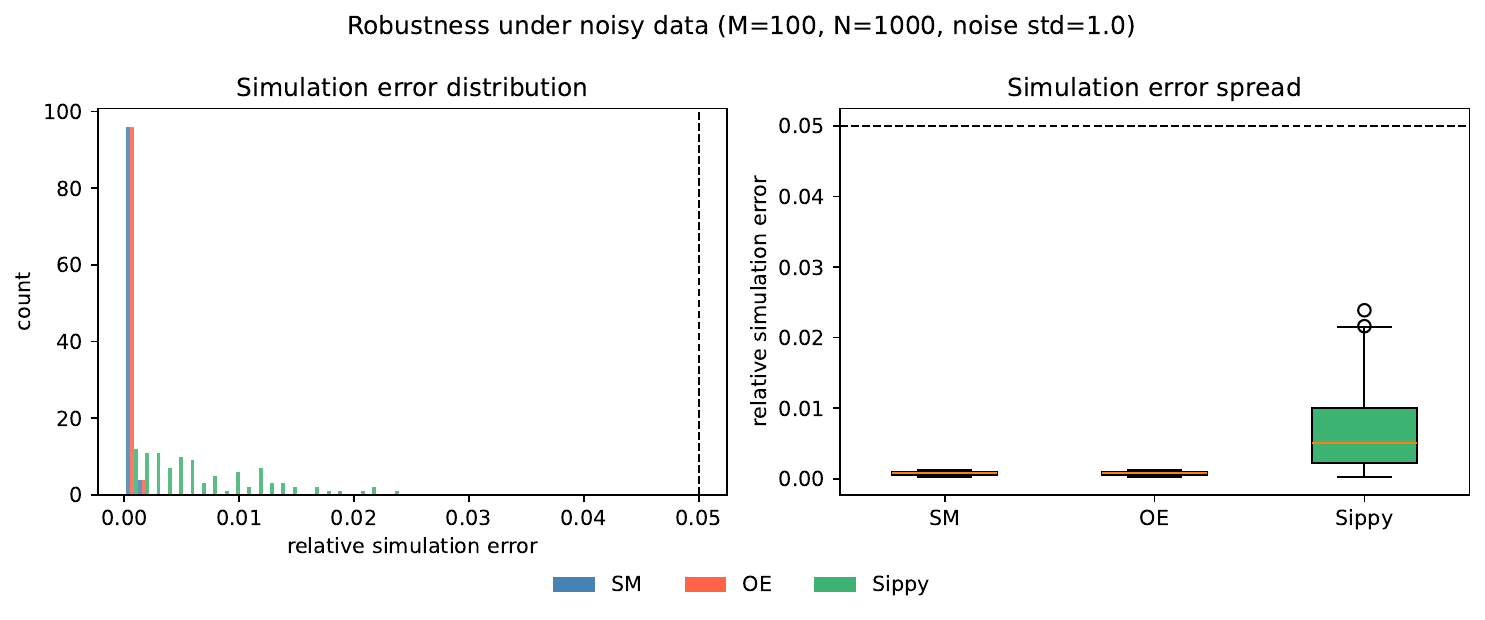}
    \caption{Relative simulation error for \texttt{pysib} SM, \texttt{pysib} OE,
    and SIPPY OE in the noisy-data comparison: histogram (left) and boxplot
    (right).}
    \label{fig:noisy-sippy-errors}
  \end{center}
\end{figure}

Table~\ref{tab:noisy-parameter} reports the empirical means and standard
deviations of the estimated OE parameters, while Table~\ref{tab:noisy-error}
and Fig.~\ref{fig:noisy-sippy-errors}
summarize the simulation errors.  Because this is a benign, near-convex
configuration, the agreement between the two \texttt{pysib} estimators is itself
informative: the Stieglitz--McBride estimator and the ARX-initialized OE PEM
estimator coincide to the reported precision and sit essentially at the
data-generating parameters, indicating that two independent search paths reach
the same minimizer.  The SIPPY OE estimates obtained in this configuration, by
contrast, exhibit a visible bias, for example, a mean \(b_1\) of \(1.020\)
against the true value \(1.0\), together with a standard deviation roughly an
order of magnitude larger and a mean simulation error about nine times larger.  The
histogram and boxplot show the same effect from the model-output perspective:
the \texttt{pysib} error distributions are tighter and closer to zero.  All
methods satisfy the success criterion in all Monte Carlo runs, so the
distinction in this moderate-noise case is not whether the plant can be
identified, but how tightly the underlying optimization returns estimates near
the known data-generating parameters.

\subsection{Filtered Continuation on a Nonconvex OE Problem}
\label{subsec:filtered-experiment}

The second experiment evaluates the filtered-continuation mechanism described in
Section~\ref{subsec:filtered-continuation-shaping}.  The plant is again
\eqref{eq:experiment-plant}, with \(M=100\) Monte Carlo runs and data length
\(N=1000\), but the additive white output noise standard deviation is increased
to \(\sigma_v=30\).  This much larger noise level is used to create a more
challenging nonconvex optimization problem, in which the standard OE search is
more likely to converge to an undesirable local minimum.  The input is again the
multisine in \eqref{eq:noisy-input}.
The comparison is between the standard \texttt{pysib} OE estimator and the
filtered OE continuation.  Both methods are evaluated by the same final
simulation-error metric \eqref{eq:simulation-error}; this is important because
the filters are used only to guide the optimization path, not to change the
final model-evaluation criterion.

\begin{table}[ht]
  \caption{Simulation-error statistics for the filtered-continuation experiment.}
  \label{tab:filtered-error}
  \centering
  \scriptsize
  \begin{tabular}{lccc}
    \toprule
    Method & Mean error & Std error & Success \\
    \midrule
    OE & \(5.69\times 10^{-2}\) & \(4.37\times 10^{-2}\) & 60\% \\
    OE filtered & \(2.17\times 10^{-2}\) & \(7.94\times 10^{-3}\) & 100\% \\
    \bottomrule
  \end{tabular}
\end{table}

\begin{figure}[htb]
  \begin{center}
    \includegraphics[width=\columnwidth]{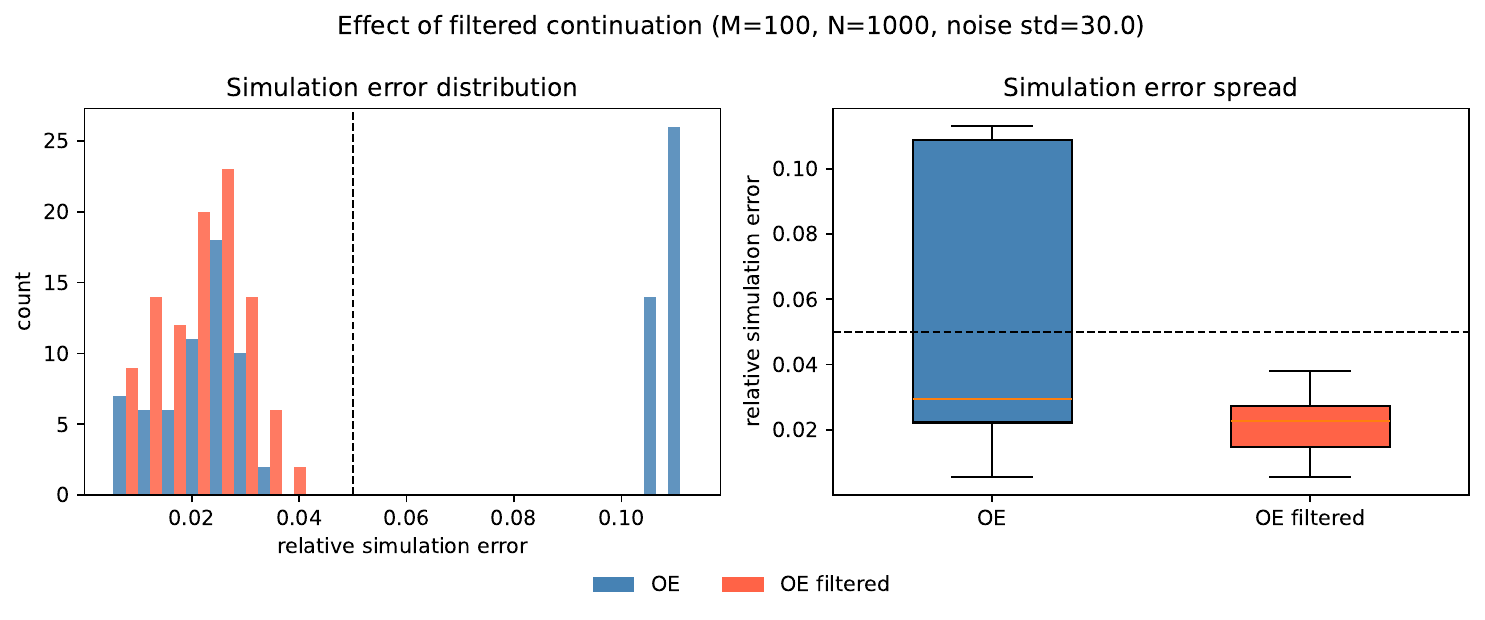}
    \caption{Relative simulation error for standard OE and filtered OE in the
    nonconvex benchmark: histogram (left) and boxplot (right).}
    \label{fig:filtered-errors}
  \end{center}
\end{figure}

The results in Table~\ref{tab:filtered-error} and
Fig.~\ref{fig:filtered-errors}
show a clear empirical difference between the two
procedures in this benchmark.  Standard OE attains the success criterion in
60\% of the runs, whereas the filtered continuation attains it in all runs and
also reduces the mean relative simulation error.  The histogram illustrates that
the improvement is not only an average effect: the filtered continuation
substantially reduces the high-error tail observed for standard OE in this
experiment.  This supports
the role of filtered continuation as a cost-shaping mechanism for this OE
problem.  The conclusion is deliberately empirical: the experiment
demonstrates improved behavior in the tested SISO OE setting, but it does not
imply a global optimality guarantee for nonconvex PEM.

\section{Conclusion}
\label{sec:conclusion}

This article presented \texttt{pysib}, an open-source Python toolbox for SISO
polynomial system identification.  Three aspects were described: a dedicated
numerical strategy for nonlinear PEM estimation, the software implementation
that exposes this strategy through a compact public interface, and the numerical
experiments used to examine the behavior of the toolbox.

The numerical strategy combines several complementary elements.  An auxiliary
ARX estimate supplies a deterministic initial point with dimensions compatible
with the requested nonlinear structure.  A smoothed-gradient phase then provides
a conservative initial decrease of the PEM criterion, before an incremental
Gauss--Newton phase introduces curvature information for local refinement.  The
filtered variants add a separate continuation mechanism: instead of changing
only the step taken by the optimizer, they change the sequence of finite-sample
criteria presented to it and then return to the original unfiltered PEM problem.

The software contribution is to expose this strategy through a compact Python
interface while keeping a consistent polynomial model representation across
estimators.  The returned pair \((\hat\theta,m)\) separates parameter inspection
from model evaluation, and the same \texttt{predict} and \texttt{simulate}
routines can be applied to models produced by different estimators.  Together
with the PyPI release, public repository, archived DOI, experiment scripts, and
automated tests, this interface is intended to make the numerical results
reproducible rather than dependent on undocumented implementation details.

The numerical evidence focused on OE benchmarks.  In the noisy-data comparison,
the \texttt{pysib} SM and OE estimates were tightly concentrated around the true
parameters and attained a small mean simulation error.  In the nonconvex OE
experiment, filtered continuation improved the empirical success rate from 60\%
for standard OE to 100\% for filtered OE, supporting its role as a cost-shaping
mechanism in that setting.  These results should be read as reproducible
evidence for the tested SISO OE problems, not as a general performance claim for
all polynomial identification settings.

\section*{Future Work}

Several directions follow naturally from the current scope.  The most immediate
one is a systematic experimental evaluation of the ARMAX and BJ routines, already
implemented with the same optimization strategy described in
Section~\ref{sec:optimization-initialization}.  A longer-term extension is to
MIMO polynomial input--output models, where the five-polynomial convention and
the common prediction and simulation interface could be generalized to
multi-variable structures.  Further work on filtered continuation could explore
adaptive filter schedules or filter families beyond fixed first-order low-pass
designs.  Finally, broader comparisons with other Python
identification packages and with the MATLAB System Identification Toolbox would
help position \texttt{pysib} in the wider ecosystem.

\section*{Declaration of generative AI and AI-assisted technologies in the manuscript preparation process}

During the preparation of this work the author used Claude (Anthropic), GPT
(OpenAI), and DeepSeek in order to improve the clarity and language of the
manuscript and to assist with software development.  After using these
tools/services, the author reviewed and edited the content as needed and takes
full responsibility for the content of the published article.

\bibliographystyle{ifacconf}
\bibliography{references}

\end{document}